# Crawling on directional surfaces

P. Gidoni, G. Noselli and A. DeSimone[*]

*SISSA, International School for Advanced Studies, Trieste, Italy.*

January 23, 2014

**Abstract**

In this paper we study crawling locomotion based on directional frictional interactions, namely, frictional forces that are sensitive to the sign of the sliding velocity. Surface interactions of this type are common in biology, where they arise from the presence of inclined hairs or scales at the crawler/substrate interface, leading to low resistance when sliding 'along the grain', and high resistance when sliding 'against the grain'. This asymmetry can be exploited for locomotion, in a way analogous to what is done in cross-country skiing (classic style, diagonal stride).

We focus on a model system, namely, a continuous one-dimensional crawler and provide a detailed study of the motion resulting from several strategies of shape change. In particular, we provide explicit formulae for the displacements attainable with reciprocal extensions and contractions (breathing), or through the propagation of extension or contraction waves. We believe that our results will prove particularly helpful for the study of biological crawling motility and for the design of bio-mimetic crawling robots.

## 1 Introduction

The study of locomotion of biological organisms and bio-mimetic engineered replicas is receiving considerable and increasing attention in the recent literature [1, 6, 7, 11, 16–18, 21, 26, 33]. In several cases, such as motility at the micron scale accomplished by unicellular organisms, or such as the ability to navigate on rough terrains exhibited by insects, worms, snakes, etc., Nature has elaborated strategies that surpass those achievable through current engineering design. The combination of quantitative observations, theoretical and computational modelling, design and optimization of bio-inspired artefacts is however leading to fast progress both in the understanding of the options Nature has selected and optimized through evolution, and on the possibility of replicating them (or even improving upon them) in man-made devices.

For example, the swimming strategies of unicellular organisms can be understood, starting from videos of their motion captured with a microscope and processed with machine-learning techniques [6], by using tools from geometric control theory [2, 4]. In fact, self-propulsion at low Reynolds numbers [30] arises from non-reciprocal looping in the space of shape parameters [2, 6], it can be replicated by using actuation strategies that can induce non-reciprocal shape changes [5, 18], and optimized by solving optimal control problems [2, 4].

---

[*]Corresponding author: `desimone@sissa.it`

Crawling motility on solid substrates of some model organisms (snails, earthworms, etc.) can be understood using similar techniques. In the case of crawlers exploiting dry friction, or lubricating fluid layers with complex rheology (such as the mucus secreted by snails [11, 13]), resistance forces are nonlinear functions of the sliding velocity and locomotion is typically accomplished through stick-and-slip. Even when resistance forces are linear in the sliding velocity, if they also depend on the size of the contact region, then locomotion is still possible, provided that more elaborate strategies are employed [16, 17, 28]. These are very similar to those that are effective in low Reynolds number swimming, and show that the transition between crawling and swimming motility is much more blurred than what was previously thought.

The results above may provide a useful theoretical framework on the way of a more detailed understanding of crawling motility of metastatic tumor cells, neuronal growth cones etc., see, e.g., [10, 21]. In addition, they may provide valuable new concepts in applications, by helping the practical design of a new generation of soft bio-inspired robots ranging from crawlers able to advance on rough terrains, to microscopic devices that may navigate inside the human body for diagnostic or therapeutic purposes [7, 18, 23, 35].

Much of the physics of the problem is contained in the question of which are the minimal mechanisms needed to make (efficient) self-propulsion possible. Here we concentrate on the question on how is it possible to extract positional change (i.e., a non-periodic history of positions) from reciprocal shape changes (i.e., a very restrictive class of periodic histories of shape change, obtained by tracing backward and forward an open curve in shape space). The famous 'scallop theorem' is precisely the statement that this is impossible for low Reynolds number swimming, see [30]. In addition, we study the motion produced by the propagation of traveling waves of contraction or extension, which is a typical strategy for self-propulsion in biology.

The interest for reciprocal shape changes arises from the fact that they can be easily accomplished by natural or artificial actuation: the breathing motion of a balloon (or of a bio-membrane) inflated and deflated by cyclic variations of (osmotic) pressure, or the motion of a specimen of a stimulus-responsive material (e.g., a shape-memory alloy) under cyclic actuation (e.g., temperature change) are all relevant examples. The conditions under which such oscillatory motions can be rectified to produce non zero net displacements has been the object of several studies, see, e.g., [12, 16, 22, 27]. In this paper, we analyse quasi-static crawling in the presence of 'directional' interactions with the environment and study in detail a model of continuous one-dimensional crawlers on directional surfaces. By this, we mean a situation in which the resistance force is not odd in the velocity: this may arise, for instance, when the substrate is hairy or it is shaped as a ratchet, or else when the interaction with the substrate is mediated by oblique flexible filaments or bristles (so that, if one reverses the sign of the velocity and moves against the grain, then the resistance force does not only change in sign, but may also change in magnitude). Concrete examples of such biological or bio-inspired directional surfaces are reviewed, e.g., in [25]. Prototypes of micro-robots exploring this motility strategy are presently being manufactured and tested, and will be described in a forthcoming paper [29].

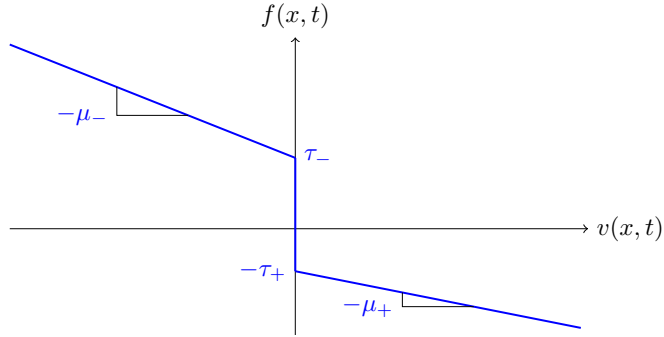

Figure 1: The general force-velocity law (2.1) for friction used in this study.

## 2 Formulation of the problem

In this study we generalize the approach to quasi-static crawling introduced in [16, 17, 28] to the case of directional substrates, namely, substrates on which the resistance to motion is sensitive to the sign of the sliding velocity. More precisely, we consider cases in which the mechanical interactions between the crawler and the substrate on which it moves are described by a force per unit (current) length denoted by $f(x,t)$, which we call 'friction force'. By *directional substrate* we mean a surface such that the friction exerted on the crawler at one point depends (only) on the velocity at that point according to a force-velocity law that is not odd in the velocity. A relevant example is the following one-dimensional force-velocity law of *Bingham*-type

$$f(x,t) = \begin{cases} \tau_- - \mu_- v(x,t) & \text{if } v(x,t) < 0, \\ \tau \in [-\tau_+, \tau_-] & \text{if } v(x,t) = 0, \\ -\tau_+ - \mu_+ v(x,t) & \text{if } v(x,t) > 0, \end{cases} \quad (2.1)$$

where $\tau_-, \tau_+, \mu_-, \mu_+$ are all non-negative material parameters[1], see Fig. 1.

There are two interesting special cases of (2.1), obtained by setting either $\mu_+ = \mu_- = 0$, or $\tau_+ = \tau_- = 0$. We refer to them as the *dry friction* and the *Newtonian friction* case, respectively, because they are reminiscent of the tangential forces arising either from dry friction, or from the drag due to a Newtonian viscous fluid, see Fig. 2. In the case of dry friction, the force depends only on the sign of the velocity ($\mu_+ = \mu_- = 0$), whereas in the Newtonian case there are no yield forces ($\tau_+ = \tau_- = 0$), so that friction depends linearly on speed through a coefficient determined by the direction of motion.

We study a straight, one-dimensional crawler moving along a straight line. Let the coordinate $X$ describe the crawler's body in the reference configuration. The left end of the body is denoted with $X_1 = 0$, while the right end with $X_2 = L$, where $L$ is the reference length. The motion of the crawler is described by the function

$$x(X,t) = x_1(t) + s(X,t), \quad (2.2)$$

where $x_1(t) = x(X_1, t)$ is the current position of the left end of the crawler (similarly, we define $x_2(t) = x(X_2, t)$ as the current position of the right end), while the arc-length $s(X,t)$, which is the current distance of point $X$ from the left end, describes its shape in the deformed configuration. By definition we have $s(0,t) = 0$, while, denoting with a prime the derivative with respect to $X$, we guarantee that the deformation described by (2.2) is one-to-one for every $t$ by assuming that

$$s'(X,t) > 0 \,. \tag{2.3}$$

---
[1]We exclude the trivial case when all the parameters vanish ($\mu_+ = \mu_- = \tau_+ = \tau_- = 0$) and therefore no frictional interaction with the substrate occurs.

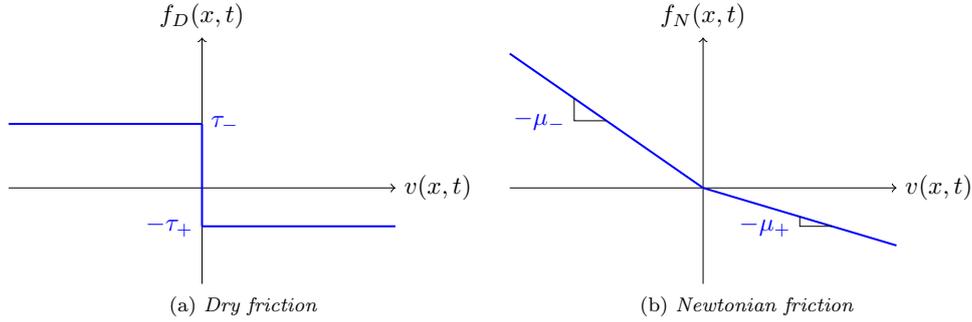

Figure 2: Two special cases of the force-velocity law (2.1) of Fig. 1.

The length $l(t)$ of the crawler at time $t$ is given by

$$l(t) = \int_0^L s'(X,t) \, \mathrm{d}X, \tag{2.4}$$

and the Eulerian velocity $v(x,t)$ at position $x$ of the crawler and time $t$ reads

$$v(x,t) = \dot{x}(X_x, t) = \dot{x}_1(t) + \dot{s}(X_x, t), \tag{2.5}$$

where $X_x = s^{-1}(x - x_1(t), t)$.

We assume that the crawler is able to control its shape, namely, to freely prescribe $s(X,t)$ subject only to the constraint (2.3). Moreover, we neglect inertia and make use of the force balance

$$F(t) = \int_0^{l(t)} f(x_1(t) + s, t) \, \mathrm{d}s = 0 \tag{2.6}$$

to obtain the velocity $\dot{x}_1(t)$ at the left hand side of the crawler.

## 3 Crawling with two shape parameters

In this section, we restrict our study to the case of a model crawler composed by two segments, namely, $\overline{X_1 X^*}$ and $\overline{X^* X_2}$, each of which is allowed to deform only affinely. Therefore, the shape of the crawler can be described by just two parameters, such as the current lengths of the two segments $l_1(t) = x^*(t) - x_1(t)$ and $l_2(t) = x_2(t) - x^*(t)$, where $x^*(t) = x(X^*, t)$. We shall consider in the following two special cases of these systems, particularly relevant to crawling on directional surfaces.

## 3.1 Crawling with only one shape parameter: breathers

We start by considering a simpler crawler made of a single segment that can only deform affinely, so that $s(X, t)$ can be expressed as a function of the current length $l(t)$ in the following way

$$s(X, t) = \frac{X}{L} l(t). \tag{3.1}$$

This model can also be obtained as a special case of the two-segment crawler subject to the additional constraint

$$\frac{\dot{l}_1(t)}{l_1(t)} = \frac{\dot{l}_2(t)}{l_2(t)}. \tag{3.2}$$

By making use of equations (2.5) and (3.1), the velocity is obtained as

$$v(x_1(t) + s, t) = \dot{x}_1(t) + \frac{s}{l(t)} \dot{l}(t), \tag{3.3}$$

a linear function of the arc-length $s \in [0, l(t)]$ vanishing at one point at most for $\dot{l}(t) \neq 0$. This implies that the force balance can be satisfied only if the velocity (and hence the force) assumes different signs along the crawler. More precisely, we argue from (3.3) that:

- if $\dot{l}(t) > 0$ (elongation), then $\dot{x}_1 < 0$ and $\dot{x}_2 > 0$;
- if $\dot{l}(t) < 0$ (contraction), then $\dot{x}_1 > 0$ and $\dot{x}_2 < 0$.

We conclude that the two ends of the crawler always move in opposite directions, and there exists $\bar{s}(t) \in (0, l(t))$ such that $v(\bar{s}(t), t) = 0$. By equation (3.3) we get

$$\bar{s}(t) = -\frac{\dot{x}_1(t) l(t)}{\dot{l}(t)}. \tag{3.4}$$

To justify the interest of one-segment crawlers (breathers) on directional surfaces, let us briefly consider the case of interactions which are odd function of the velocity, i.e., interactions such that $f(-v) = -f(v)$. By using the force balance (2.6) and equation (3.3), it can be easily shown that $\bar{s}(t) = l(t)/2$. From equation (3.4) we get $\dot{x}_1(t) = -\dot{l}(t)/2$, and thereby breathing deformation modes always lead to zero net displacement, when performed on homogeneous surfaces which are not directional.

Consider now the directional law of equation (2.1): the frictional force acting at one point of the crawler depends on whether its distance from $x_1$ is smaller or larger than $\bar{s}$. For convenience, we establish that the parameters $(\tau_1, \mu_1)$ describe the forces acting on the left side of $x_1 + \bar{s}$, and $(\tau_2, \mu_2)$ those acting on the right side. Explicitly, we set

$$(\tau_1, \mu_1) = \begin{cases} (\tau_-, \mu_-) & \text{if } \dot{l} > 0, \\ (-\tau_+, \mu_+) & \text{if } \dot{l} < 0, \end{cases} \quad \text{and} \quad (\tau_2, \mu_2) = \begin{cases} (-\tau_+, \mu_+) & \text{if } \dot{l} > 0, \\ (\tau_-, \mu_-) & \text{if } \dot{l} < 0, \end{cases} \tag{3.5}$$

so that the total force acting on the crawler reads

$$F(t) = \int_0^{\bar{s}} \left[ \tau_1 - \mu_1 \left( \dot{x}_1 + \frac{s \dot{l}}{l} \right) \right] ds + \int_{\bar{s}}^l \left[ \tau_2 - \mu_2 \left( \dot{x}_1 + \frac{s \dot{l}}{l} \right) \right] ds =$$

$$= (\mu_2 - \mu_1) \bar{s} \left( \dot{x}_1 + \frac{\bar{s} \dot{l}}{2l} \right) - \mu_2 l \left( \dot{x}_1 + \frac{\dot{l}}{2} \right) - (\tau_2 - \tau_1) \bar{s} + \tau_2 l. \tag{3.6}$$

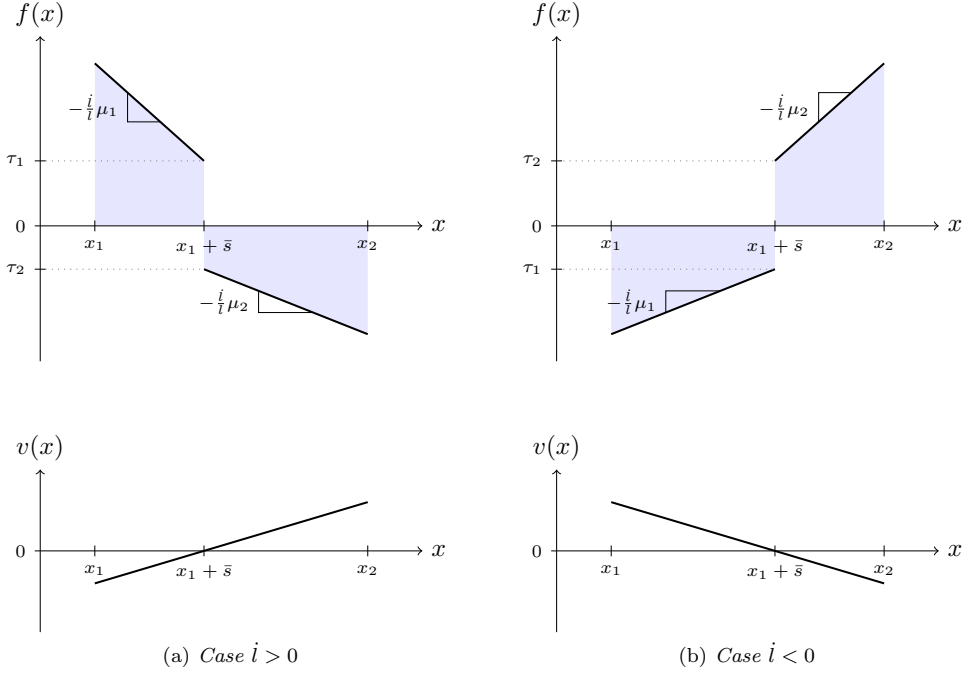

(a) Case $\dot{l} > 0$

(b) Case $\dot{l} < 0$

Figure 3: The velocity $v(x)$ and the force per unit current length $f(x)$ along a breather in the extensile and contractive case.

Replacing expression (3.4) for $\bar{s}$ and dividing by $l$, the force balance $F(t) = 0$ leads to the following equation for $\dot{x}_1$

$$\frac{(\mu_1 - \mu_2)}{2\dot{l}} \dot{x}_1^2 + \left(\frac{\tau_2 - \tau_1}{\dot{l}} - \mu_2\right) \dot{x}_1 - \frac{\mu_2 \dot{l}}{2} + \tau_2 = 0. \tag{3.7}$$

In the special case of $\mu_1 = \mu_2 = \mu$, equation (3.7) becomes linear and its solution reads

$$\dot{x}_1 = -\left(1 + \frac{\tau_1 + \tau_2}{\tau_2 - \tau_1 - \mu \dot{l}}\right) \frac{\dot{l}}{2}, \tag{3.8}$$

whereas, for $\mu_1 \neq \mu_2$, equation (3.7) is quadratic with discriminant

$$\Delta = \mu_1 \mu_2 + \frac{(\tau_2 - \tau_1)^2}{\dot{l}^2} + \frac{2}{\dot{l}}(\mu_2 \tau_1 - \mu_1 \tau_2). \tag{3.9}$$

The first two terms of the RHS of (3.9) are both nonnegative and, having excluded the null friction case, at least one of them is positive. The two parameters $\tau_1$ and $\tau_2$, when non zero, have respectively the same and the opposite sign of $\dot{l}$, so also the third term is nonnegative and equation (3.7) has two distinct real solutions

$$\dot{x}_1^{\pm} = \frac{\mu_2 + \frac{\tau_1 - \tau_2}{\dot{l}} \pm \sqrt{\Delta}}{\mu_1 - \mu_2} \dot{l} = C^{\pm} \dot{l}. \tag{3.10}$$

Table 1: Expressions of $\dot{x}_1(t)$ for the one-segment crawler in the extensile and contractive case.

| | $\mu_- = \mu_+ = \mu$, $\tau_- \geq \tau_+$ | $\mu_- > \mu_+$ |
|---|---|---|
| $\dot{l}(t) > 0:$ | $\left(\dfrac{\tau_- - \tau_+}{\tau_- + \tau_+ + \mu\|\dot{l}\|} - 1\right)\dfrac{\|\dot{l}\|}{2}$ | $\dfrac{\mu_+ + \frac{\tau_-+\tau_+}{\|\dot{l}\|} - \sqrt{\mu_-\mu_+ + \frac{(\tau_-+\tau_+)^2}{\dot{l}^2} + \frac{2}{\|\dot{l}\|}(\mu_-\tau_+ + \mu_+\tau_-)}}{\mu_- - \mu_+}\|\dot{l}\|$ |
| $\dot{l}(t) < 0:$ | $\left(\dfrac{\tau_- - \tau_+}{\tau_- + \tau_+ + \mu\|\dot{l}\|} + 1\right)\dfrac{\|\dot{l}\|}{2}$ | $\dfrac{\mu_- + \frac{\tau_-+\tau_+}{\|\dot{l}\|} - \sqrt{\mu_-\mu_+ + \frac{(\tau_-+\tau_+)^2}{\dot{l}^2} + \frac{2}{\|\dot{l}\|}(\mu_-\tau_+ + \mu_+\tau_-)}}{\mu_- - \mu_+}\|\dot{l}\|$ |

In view of (3.4), however, any admissible solution must satisfy

$$C^\pm = \frac{\dot{x}_1^\pm}{\dot{l}} = -\frac{\bar{s}}{l} \in (-1, 0), \qquad (3.11)$$

and we claim that, for any choice of the parameters, this condition is satisfied only by the solution $\dot{x}_1^-$. We start by observing that the following estimate holds for $\Delta$

$$\left(\min\{\mu_1, \mu_2\} + \frac{\tau_1 - \tau_2}{l}\right)^2 < \Delta < \left(\max\{\mu_1, \mu_2\} + \frac{\tau_1 - \tau_2}{l}\right)^2. \qquad (3.12)$$

If $\mu_1 < \mu_2$, then by (3.12) we see that both $C^-$ and $C^+$ are negative, but $C^- > -1$ while $C^+ < -1$. On the other hand, if $\mu_1 > \mu_2$, applying again (3.12), we have $C^- \in (-1, 0)$ while $C^+ > 0$.

We can assume, without loss of generality, that

$$\mu_- > \mu_+, \text{ if } \mu_- \neq \mu_+, \quad \text{or that} \quad \tau_- \geq \tau_+, \text{ if } \mu_- = \mu_+. \qquad (3.13)$$

Indeed, this amount to fixing the orientation of the $x$ axis so that the positive direction is the one of least frictional resistance, in the sense specfied by (3.13). The expressions for the velocity $\dot{x}_1(t)$ as a function of the rate of shape change are presented in Table 1.

It is interesting to notice that the velocity $\dot{x}_1(t)$ is invariant if we multiply the force-velocity law (2.1) by a positive factor. Furthermore, $\dot{x}_1(t)$ does not depend explicitly on the length $l(t)$, but just on its time derivative $\dot{l}(t)$.

It is also interesting to remark that, for the special cases of dry friction ($\mu_- = \mu_+ = 0$) and Newtonian friction ($\tau_- = \tau_+ = 0$), $\dot{x}_1(t)$ becomes linear in $|\dot{l}|$. Hence, in these situations, the displacement produced by any history of shape changes depends on the path traced in the configuration space, but not on the speed at which it is executed. In particular, the displacement produced in a cycle composed of a monotone elongation (resp. contraction) followed by a monotone return to the initial length is a linear function of the length increase (resp. decrease), through a coefficient determined by the force-velocity laws. We now examine these cases in more detail.

**Dry friction** To analyze the case of dry friction, we introduce the dimensionless parameter $\alpha = \tau_-/(\tau_- + \tau_+) \in (0, 1)$, such that the orientation assumption $\tau_- \geq \tau_+$ implies $\alpha \geq 1/2$ and the formula for the velocity $\dot{x}_1(t)$ reads

$$\dot{x}_1(t) = \begin{cases} -(1-\alpha)\dot{l}(t) < 0 & \text{if } \dot{l}(t) > 0 \text{ (elongation)}, \\ -\alpha\dot{l}(t) > 0 & \text{if } \dot{l}(t) < 0 \text{ (contraction)}. \end{cases} \quad (3.14)$$

The net displacement of the single-segment crawler, arising from a $T$-periodic shape change, can be computed by integration of equation (3.14) upon definition of $l(t)$ for $t \in [0, T]$. Let us consider the following example. The length of the crawler first increases (resp. decreases) monotonically from $L$ to $L + \delta$, and then decreases (resp. increases) from $L + \delta$ to $L$, with $\delta$ a positive (resp. negative) quantity. An example of such $T$-periodic shape function is given by $l(t) = L + \delta \sin^2(\pi t/T)$, and the net advancement after one stretching cycle simply follows as

$$\Delta_D x_1 = (2\alpha - 1)|\delta|. \quad (3.15)$$

This is proportional to the peak extension (or contraction) $\delta$ experienced by the crawler, through the non-negative coefficient $2\alpha - 1 < 1$. This coefficient approaches 1 when $\alpha$ tends to 1, and this occurs in the case of infinite contrast between the frictional resistances in the easy and hard directions ($\tau_-/\tau_+ \to \infty$). In this idealised case there is no back-sliding, and all the available extension/contraction of the crawler's body is converted into 'useful' displacement, as it is commonly assumed in the classical literature (see e.g., [1, 31]). We finally remark that in the limiting case of $\tau_- = \tau_+$ the net advancement $\Delta_D x_1$ vanishes as $\alpha = 1/2$.

**Newtonian friction** For the analysis of the Newtonian case, we introduce the dimensionless parameter $\beta = \sqrt{\mu_-/\mu_+} \in (0, +\infty)$ (note that we restrict to $\beta > 1$ in view of the orientation assumption on the $x$ axis, $\mu_- > \mu_+$). Setting $\tau_- = \tau_+ = 0$ in Table 1 we obtain

$$\dot{x}_1(t) = \begin{cases} -\dfrac{1}{\beta + 1}\dot{l}(t) < 0 & \text{if } \dot{l}(t) > 0 \text{ (elongation)}, \\ -\dfrac{\beta}{\beta + 1}\dot{l}(t) > 0 & \text{if } \dot{l}(t) < 0 \text{ (contraction)}. \end{cases} \quad (3.16)$$

We now consider the time-periodic shape change previously assumed for the case of dry friction, that is, a monotone expansion-contraction between lengths $L$ and $L + \delta$. The net displacement after one period reads

$$\Delta_N x_1 = \left(-\frac{1}{\beta + 1} + \frac{\beta}{\beta + 1}\right)|\delta| = \frac{\beta - 1}{\beta + 1}|\delta|. \quad (3.17)$$

This is again proportional to the maximum change in length $\delta$ experienced by the crawler through the positive coefficient $(\beta - 1)/(\beta + 1) \in (0, 1)$. The limiting case of infinite contrast between the frictional resistances in the easy and hard directions ($\tau_-/\tau_+ \to \infty$, and hence $\beta \to +\infty$) leads to $\Delta_N x_1 \to \delta$. Furthermore, in the limiting case of $\mu_- = \mu_+$ the net advancement $\Delta_N x_1$ vanishes, since $\beta = 1$.

### 3.2 Crawling with only one shape parameter: constant length crawlers

We turn now our attention to another special case of two-segment crawler, arising from the additional constraint of constant total length, i.e.,

$$l_1(t) + l_2(t) = L. \quad (3.18)$$

In this context, the shape of the crawler can be described by only one parameter, say, $l_1(t)$, and the arc-length $s(X,t)$ reads as

$$s(X,t) = \begin{cases} \dfrac{X}{X^*}l_1(t) & \text{if } X \in [0, X^*], \\ l_1(t) + \dfrac{L - l_1(t)}{L - X^*}(X - X^*) & \text{if } X \in (X^*, L]. \end{cases} \qquad (3.19)$$

Moreover, we have that $\dot{x}_2(t) = \dot{x}_1(t) + \dot{l}_1(t) + \dot{l}_2(t) = \dot{x}_1(t)$ and the Eulerian velocity at point $x = x_1(t) + s$ and time $t$ reads

$$v(x_1(t) + s, t) = \begin{cases} \dot{x}_1(t) + \dfrac{s}{l_1(t)}\dot{l}_1(t) & \text{if } s \in [0, l_1(t)], \\ \dot{x}_1(t) + \dfrac{L - s}{L - l_1(t)}\dot{l}_1(t) & \text{if } s \in (l_1(t), L]. \end{cases} \qquad (3.20)$$

We notice from equation (3.20) that the velocity equals zero in two points at most. As already observed for the one-segment crawler, the force balance can be satisfied only if the velocity assumes both signs along the crawler, so that there must exist two points $\bar{s}_1(t) \in (0, l_1(t))$ and $\bar{s}_2(t) \in (l_1(t), L)$ where the velocity vanishes. From equation (3.20) we conclude that

$$\bar{s}_1(t) = -\frac{\dot{x}_1(t) l_1(t)}{\dot{l}_1(t)} \quad \text{and} \quad \bar{s}_2(t) = L + \frac{\dot{x}_1(t)(L - l_1(t))}{\dot{l}_1(t)}, \qquad (3.21)$$

and we further observe that the following relation holds between $\bar{s}_1(t)$ and $\bar{s}_2(t)$

$$\bar{s}_2(t) = L - \frac{L - l_1(t)}{l_1(t)}\bar{s}_1(t). \qquad (3.22)$$

For a two-segment, constant length crawler, the velocity assumes one sign in the interval $(\bar{s}_1(t), \bar{s}_2(t))$, and the other one outside that interval. We adapt the definition of $(\tau_1, \mu_1)$ and $(\tau_2, \mu_2)$ given above by replacing $\dot{l}$ with $\dot{l}_1$ in (3.5). Thus $(\tau_2, \mu_2)$ refer to the interval $(\bar{s}_1(t), \bar{s}_2(t))$, while $(\tau_1, \mu_1)$ are the friction parameters in $[0, \bar{s}_1(t))$ and $(\bar{s}_2(t), L]$. With these positions, the total force acting on the crawler is

$$F(t) = \int_0^{\bar{s}_1}\left[\tau_1 - \mu_1\left(\dot{x}_1 + \frac{s\dot{l}_1}{l_1}\right)\right]ds + \int_{\bar{s}_1}^{l_1}\left[\tau_2 - \mu_2\left(\dot{x}_1 + \frac{s\dot{l}_1}{l_1}\right)\right]ds +$$

$$+ \int_{l_1}^{\bar{s}_2}\left[\tau_2 - \mu_2\left(\dot{x}_1 + \frac{(L-s)\dot{l}_1}{L - l_1}\right)\right]ds + \int_{\bar{s}_2}^{L}\left[\tau_1 - \mu_1\left(\dot{x}_1 + \frac{(L-s)\dot{l}_1}{L - l_1}\right)\right]ds, \qquad (3.23)$$

and, using equation (3.22), the force balance $F(t) = 0$ can be written as

$$\left(1 + \frac{L - l_1}{l_1}\right)\left\{\int_0^{\bar{s}_1}\left[\tau_1 - \mu_1\left(\dot{x}_1 + \frac{s\dot{l}_1}{l_1}\right)\right]ds + \int_{\bar{s}_1}^{l_1}\left[\tau_2 - \mu_2\left(\dot{x}_1 + \frac{s\dot{l}_1}{l_1}\right)\right]ds\right\} = 0. \qquad (3.24)$$

Comparing the last equation with (3.6), we notice that the force balance on the whole crawler is satisfied if and only if it is independently satisfied on each of the two segments, assuming no exchange of force between them (see also Fig. 4). This means that a two-segment crawler with constant length is equivalent to two adjacent but independent one-segment crawlers that are 'well coordinated' (as a consequence of the constant total length constraint): they move remaining adjacent, neither pushing nor pulling each other.

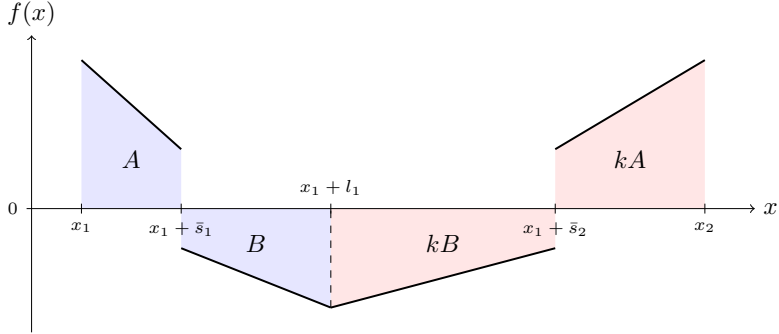

Figure 4: Graphical interpretation of equation (3.24). The two integrals in (3.24) correspond respectively to $A$ and $B$, while $k = (L - l_1)/l_1$. The force balance on the whole crawler is $(1 + k)(A + B) = 0$. It can be satisfied if and only if $A + B = 0$, that is exactly the force balance on the first segment.

It follows that the motion of $x_1$ can be obtained by applying the results for single-segment crawlers to the first segment alone. In particular, the expressions for $\dot{x}_1(t)$ of Table 1 and equations (3.14) and (3.16) hold for the two-segment crawler with constant length if we replace $\dot{l}(t)$ with $\dot{l}_1(t)$. Equations (3.15) and (3.17) also hold if we consider a periodic motion where the first segment experiences a monotone elongation-contraction between lengths $L_1$ and $L_1 + \delta$, being $L_1$ the reference length of the first segment.

## 3.3 A composite stride for a two-segment crawler

We notice that the two examples of periodic shape change considered so far, each of which exploits just one shape parameter, both produce a positive displacement, namely, a net displacement in the direction of least frictional resistance. We would like to investigate whether, by suitably composing these 'elementary' shape changes, we can obtain a net displacement in the direction of maximal frictional resistance, i.e., a negative displacement in view of our orientation assumption (3.13). We will determine below the conditions under which this 'riding against the largest friction' is indeed possible.

Given any $\delta, \lambda > 0$ and $h > 1$, we define the following points in the $(l_1, l_2)$ shape parameters space

$$A = (\lambda + \delta, \lambda), \qquad B = (\lambda, \lambda + \delta),$$
$$C = (h\lambda, h(\lambda + \delta)), \qquad D = (h(\lambda + \delta), h\lambda).$$

We shall now explore the case of shape changes arising from the closed polygonal chain with vertices $A$, $B$, $C$ and $D$, see Fig. 5.

**Dry Friction** In order to compute the net displacement arising from the closed loop depicted in Fig. 5, it is useful to first notice that:

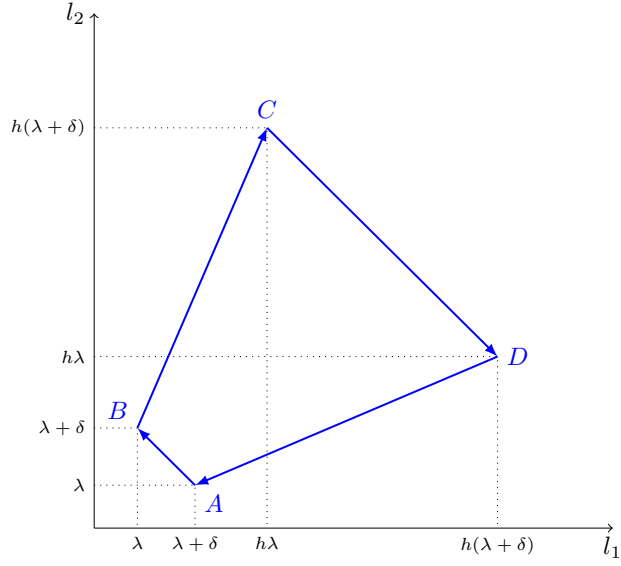

Figure 5: An example of periodic shape change for a two-parameters crawler.

- the two segments $A \to B$ and $C \to D$ keep the total length $l_1 + l_2$ of the crawler constant, so that, in view of equation (3.24), their contributions to the displacement can be evaluated by means of equation (3.14) applied to the first segment only;

- the two stride paths $B \to C$ and $D \to A$ satisfy condition (3.2), so that their 'breathing' contributions to the displacement can be evaluated by means of equation (3.14) applied to the crawler as a whole.

Specifically, the four contributions to the displacement read

$$\begin{aligned}
\Delta_D^{AB} &= \alpha\,\delta, \\
\Delta_D^{BC} &= -(1-\alpha)(h-1)(2\lambda+\delta), \\
\Delta_D^{CD} &= -(1-\alpha)h\,\delta, \\
\Delta_D^{DA} &= -\alpha(1-h)(2\lambda+\delta),
\end{aligned} \qquad (3.25)$$

and thereby the net displacement $\Delta_D x_1$ produced after one cycle is

$$\Delta_D x_1 = \alpha\left[4\lambda(h-1) + \delta(3h-1)\right] - 2\lambda(h-1) - \delta(2h-1). \qquad (3.26)$$

A negative net displacement, $\Delta_D x_1 < 0$, can therefore be obtained only if

$$2\alpha - 1 < \frac{1}{\dfrac{4\lambda}{\delta} + \dfrac{3h-1}{h-1}} < \frac{1}{3}, \qquad (3.27)$$

where the upper bound can be approached when $h \to +\infty$ and $\lambda/\delta \to 0$. It turns out that, in this limit case, a negative displacement is possible only if $\alpha < 2/3$, i.e., only if $\tau_-/\tau_+ < 2$.

**Newtonian friction** To explore the case of Newtonian friction, we proceed just as in the case of dry friction, but using equation (3.16) instead. The four contributions to the displacement read now

$$\Delta_N^{AB} = \frac{\beta}{\beta+1}\delta,$$

$$\Delta_N^{BC} = -\frac{1}{\beta+1}(h-1)(2\lambda+\delta),$$

$$\Delta_N^{CD} = -\frac{1}{\beta+1}h\,\delta, \quad (3.28)$$

$$\Delta_N^{DA} = -\frac{\beta}{\beta+1}(1-h)(2\lambda+\delta),$$

and thereby the net displacement $\Delta_N x_1$ produced after one cycle is

$$\Delta_N x_1 = \frac{\beta}{\beta+1}\left[2\lambda(h-1)+\delta h\right] - \frac{1}{\beta+1}\left[2\lambda(h-1)+\delta(2h-1)\right]. \quad (3.29)$$

We thus obtain a negative net displacement, $\Delta_N x_1 < 0$, only if

$$\beta - 1 < \frac{1}{\frac{2\lambda}{\delta}+\frac{h}{h-1}} < 1, \quad (3.30)$$

where the upper bound can be approached for $h \to +\infty$ and $\lambda/\delta \to 0$. Thus, in this limit case, a negative displacement is possible only if $\beta < 2$, i.e., only if the ratio of $\mu_-/\mu_+ < 4$.

## 4 Crawling with square waves

The purpose of this section is to extend our analyses by exploring the case of shape changes arising from extension (or contraction) traveling waves. In particular, we consider a square stretching wave of width $\delta < L$ and amplitude $\varepsilon$, traveling rightwards along the crawler with speed $c > 0$. The shape $s(X,t)$ is assumed to be $(L+\delta)/c$ periodic in the time variable $t$, and defined as follows

$$s(X,t) = \begin{cases} \left.\begin{array}{ll} X(1+\varepsilon) & \text{for } X \in [0, ct) \\ X + \varepsilon ct & \text{for } X \in [ct, L] \end{array}\right\} & \text{if } t \in [0, \delta/c), \\ \left.\begin{array}{ll} X & \text{for } X \in [0, ct-\delta) \\ X + \varepsilon(X+\delta-ct) & \text{for } X \in [ct-\delta, ct) \\ X + \varepsilon\delta & \text{for } X \in [ct, L] \end{array}\right\} & \text{if } t \in [\delta/c, L/c), \\ \left.\begin{array}{ll} X & \text{for } X \in [0, ct-\delta) \\ X + \varepsilon(X+\delta-ct) & \text{for } X \in [ct-\delta, L] \end{array}\right\} & \text{if } t \in [L/c, (L+\delta)/c), \end{cases} \quad (4.1)$$

so that the current length of the crawler reads

$$l(t) = \begin{cases} L + \varepsilon ct & \text{if } t \in [0, \delta/c), \\ L + \varepsilon\delta & \text{if } t \in [\delta/c, L/c), \\ L + \varepsilon(L+\delta-ct) & \text{if } t \in [L/c, (L+\delta)/c). \end{cases} \quad (4.2)$$

Therefore, by making use of equations (2.5) and (4.1), we obtain the Eulerian velocity at point $x = x_1(t) + s$ as

$$v(x,t) = \begin{cases} \begin{rcases} \dot{x}_1(t) & \text{for } s \in [0, (1+\varepsilon)c\,t) \\ \dot{x}_1(t) + \varepsilon\,c & \text{for } s \in [(1+\varepsilon)c\,t, L + \varepsilon\,c\,t] \end{rcases} & \text{if } t \in [0, \delta/c), \\ \begin{rcases} \dot{x}_1(t) & \text{for } s \in [0, c\,t - \delta) \\ \dot{x}_1(t) - \varepsilon\,c & \text{for } s \in [c\,t - \delta, c\,t + \varepsilon\,\delta) \\ \dot{x}_1(t) & \text{for } s \in [c\,t + \varepsilon\,\delta, L + \varepsilon\,\delta] \end{rcases} & \text{if } t \in [\delta/c, L/c), \\ \begin{rcases} \dot{x}_1(t) & \text{for } s \in [0, c\,t - \delta) \\ \dot{x}_1(t) - \varepsilon\,c & \text{for } s \in [c\,t - \delta, L + \varepsilon(L + \delta - c\,t)] \end{rcases} & \text{if } t \in [L/c, (L+\delta)/c). \end{cases} \quad (4.3)$$

We focus now our attention on the case of extension waves, such that $\varepsilon > 0$. The case of contraction waves, with $-1 < \varepsilon < 0$, can be treated similarly. From equation (4.3) we observe that, at any time $t$, the velocity along the crawler can assume only two values: a certain velocity $\nu(t)$ at the points where no deformation occurs ($s'(X,t) = 1$) and a lower velocity $\nu(t) - \varepsilon\,c$ at the points experiencing elongation ($s'(X,t) = 1 + \varepsilon$). Obviously, force balance dictates that $\nu(t) \geq 0$ for extension waves (resp. $\nu(t) \leq 0$ for contraction waves) and, in principle, two qualitatively different situations are possible: if $\nu(t) = 0$, then a stick-slip behavior takes place (with 'slipping' occurring in the elongating part, with velocity $-\varepsilon\,c$, and 'sticking' elsewhere), whereas for $\nu(t) \neq 0$ sliding occurs throughout the crawler.

Furthermore, we notice that for extension waves the velocity $\nu(t)$ is restricted to $\nu(t) \leq \varepsilon\,c$ (resp. $\nu(t) \geq \varepsilon\,c$ for contraction waves), and that the choice of $\nu(t) = \varepsilon\,c$ is compatible with the force balance only when the part of crawler being stretched is sufficiently large with respect to its total length. In fact, let us consider the time interval $t \in [0, \delta/c)$, during which the extension wave enters the crawler at its left end. For any time $t$ such that

$$c\,t < L \frac{\tau_+ + \mu_+ \varepsilon\,c}{(1+\varepsilon)\tau_- + \tau_+ + \mu_+ \varepsilon\,c}, \quad (4.4)$$

the following estimate applies to the total force $F(t)$ acting on the crawler

$$F(t) = \int_0^{(1+\varepsilon)c\,t} \tau(s,t)\,\mathrm{d}s - (L - c\,t)(\tau_+ + \mu_+ \varepsilon\,c) \leq (1+\varepsilon)\tau_- c\,t - (L - c\,t)(\tau_+ + \mu_+ \varepsilon\,c) < 0, \quad (4.5)$$

and thus the force balance does not hold. In other words, an 'inverted' stick-slip crawler, where sticking occurs along the deformed part and slipping along the other one, is in general not admissible in the context of our analysis, where the stride is given by equation (4.1). The only exception is the trivial case of $\tau_+ = \mu_+ = 0$.

In the following sections we shall assume that, for a given crawler, only one of the two modes of locomotion can be activated, and we will separately consider *stick-slip* crawlers ($\nu(t) = 0$) and *sliding* crawlers ($\nu(t) \neq 0$).

## 4.1 Stick-slip crawlers

We first explore the case of stick-slip crawlers, such that the Ansatz $\nu(t) = 0$ for every $t$ yields

$$\dot{x}_1(t) = \begin{cases} -\varepsilon\,c & \text{for } t \in [0, \delta/c), \\ 0 & \text{for } t \in [\delta/c, (L+\delta)/c), \end{cases} \quad (4.6)$$

and time integration of (4.6) in the interval $t \in [0, (L+\delta)/c)$ immediately leads to the expression

$$\Delta x_1 = -\varepsilon\,\delta \qquad (4.7)$$

for the net displacement after one period of the square wave.

We just need to check the compatibility of the Ansatz with the force balance. To this end, we compute the overall force exerted on the crawler, which, for an extension wave ($\varepsilon > 0$), reads

$$F(t) = \begin{cases} (\tau_- + \mu_-\varepsilon c)(1+\varepsilon)ct + \int_{(1+\varepsilon)ct}^{L+\varepsilon ct} \tau(s,t)\,\mathrm{d}s & \text{for } t \in [0, \delta/c), \\ \int_0^{ct-\delta} \tau(s,t)\,\mathrm{d}s + (\tau_- + \mu_-\varepsilon c)(1+\varepsilon)\delta + \int_{ct+\varepsilon\delta}^{L+\varepsilon\delta} \tau(s,t)\,\mathrm{d}s & \text{for } t \in [\delta/c, L/c), \\ \int_0^{ct-\delta} \tau(s,t)\,\mathrm{d}s + (\tau_- + \mu_-\varepsilon c)(1+\varepsilon)(L-ct+\delta) & \text{for } t \in [L/c, (L+\delta)/c), \end{cases} \qquad (4.8)$$

and we further notice that the most restrictive condition to $F(t) = 0$ is given by the middle term of (4.8), which specifically requires

$$\delta \leq \frac{\tau_+ L}{(\tau_- + \mu_-\varepsilon\,c)(1+\varepsilon) + \tau_+} = \delta_+^{\max}. \qquad (4.9)$$

The case of contraction waves ($-1 < \varepsilon < 0$) can be studied similarly and leads to the following restriction on $\delta$

$$\delta \leq \frac{\tau_- L}{(\tau_+ - \mu_+\varepsilon\,c)(1+\varepsilon) + \tau_-} = \delta_-^{\max}. \qquad (4.10)$$

We notice that $\tau_+ = 0$ implies $\delta_+^{\max} = 0$ and, likewise, $\tau_- = 0$ implies $\delta_-^{\max} = 0$. Therefore, no stick-slip behaviour can occur on a Newtonian substrate and the largest achievable displacement, at fixed $\varepsilon$, is given by $-\varepsilon\,\delta_\pm^{\max}$.

The displacement (4.7) produced by an elongation wave spanning the crawler's body once is always negative (opposite to the wave direction), whereas it is always positive (concordant to the wave direction) for a contraction wave. Furthermore, we observe that the net advancement $\Delta x_1 \to \delta$ in the limit $\varepsilon \to -1$, and, as we will see, this is a feature in common with sliding crawlers. This is due to the fact that, as $\varepsilon \to -1$, the portion of crawler experiencing deformation collapses to a single point: the force balance is then trivially satisfied, with no friction being exerted, and the resulting motion is determined exclusively by geometrical reasons, rather than dynamical ones.

**Dry friction** The coefficients $\mu_+$ and $\mu_-$ play only a minor role in stick-slip crawling. In fact, they only reduce the set of admissible square waves, see equations (4.9)-(4.10), and hence the maximum achievable displacement. Thus, dry friction is an ideal environment to study stick-slip behaviour. By making use of (4.7) and equations (4.9)-(4.10), the maximum achievable advancement for a traveling wave of fixed $\varepsilon$ is obtained as

$$\Delta_D x_1 = \begin{cases} -\dfrac{\varepsilon(1-\alpha)}{1+\varepsilon\alpha} L & \text{for } \varepsilon > 0 \text{ (extension wave)}, \\ -\dfrac{\varepsilon\alpha}{1+\varepsilon(1-\alpha)} L & \text{for } -1 < \varepsilon < 0 \text{ (contraction wave)}. \end{cases} \qquad (4.11)$$

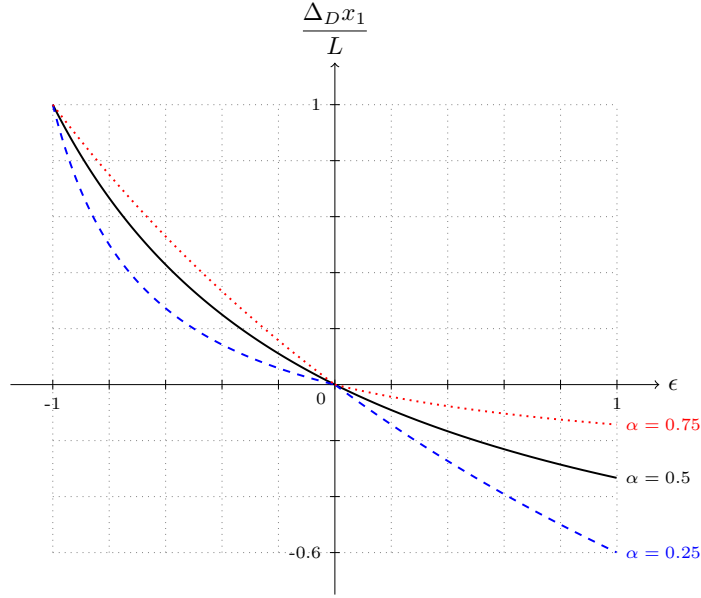

Figure 6: Maximum displacement $\Delta_D x_1/L$ for a stick-slip crawler in the case of dry friction as a function of $\varepsilon \in (-1, 1)$ and for $\alpha = \{0.25, 0.5, 0.75\}$.

The maximum displacement after one stretching cycle is shown in Fig. 6 in dimensionless form as a function of $\varepsilon$ and for $\alpha = \{0.25, 0.5, 0.75\}$. The net advancement is always negative (positive) for extension (resp. contraction) waves, and its magnitude decreases (resp. increases) as $\alpha$ is increased.

## 4.2 Sliding crawlers

We turn now to the case of sliding crawlers, so that no stick-slip behaviour can occur and $\nu(t) > 0$ for $t \neq n(L + \delta)/c$ with $n$ integer (at these times no deformation takes place and the velocity vanishes everywhere along the crawler). Specifically, we consider extensional waves ($\varepsilon > 0$) and study separately the force balance equation in the following three stages of deformation.

*Stage a:* $t \in (0, \delta/c)$. During this first time interval the square wave enters the crawler at its left end. As already discussed in section 4, force balance requires that $0 < \nu(t) < \varepsilon c$ and hence the velocity of the left end of the crawler is restricted to $-\varepsilon c < \dot{x}_1(t) < 0$. Thus, the force balance equation becomes

$$[\tau_- - \mu_- \dot{x}_1(t)](1+\varepsilon)ct - [\tau_+ + \mu_+(\dot{x}_1(t) + \varepsilon c)](L - ct) = 0, \qquad (4.12)$$

from which we get

$$\dot{x}_1(t) = \frac{\tau_-(1+\varepsilon)ct - (\tau_+ + \mu_+ \varepsilon c)(L - ct)}{\mu_-(1+\varepsilon)ct + \mu_+(L - ct)}. \qquad (4.13)$$

Taking into account the restrictions upon $\dot{x}_1(t)$, the solution (4.13) is admissible only if

$$\tau_+ = 0 \quad \text{and} \quad \delta < \frac{\mu_+ \varepsilon c}{\mu_+ \varepsilon c + \tau_-(1+\varepsilon)} L, \qquad (4.14)$$

and we notice that this implies $\mu_+ \neq 0$, for else $\delta = 0$ and no motion occurs.

Hereafter we assume[2] that $(1+\varepsilon)\mu_- \neq \mu_+$, and integration of (4.13) in the time interval $t \in (0, \delta/c)$ immediately provides the expression of the first contribution to the displacement, namely,

$$\Delta x_1^a = \frac{\delta[(1+\varepsilon)\tau_- + \mu_+\varepsilon c]}{c[(1+\varepsilon)\mu_- - \mu_+]} + \frac{L(1+\varepsilon)(\tau_- + \mu_-\varepsilon c)\mu_+}{c[(1+\varepsilon)\mu_- - \mu_+]^2} \ln\left[\frac{L\mu_+}{\delta(1+\varepsilon)\mu_- + (L-\delta)\mu_+}\right]. \quad (4.15)$$

*Stage b:* $t \in [\delta/c, L/c)$. At any instant of this interval, the square wave of width $\delta$ is entirely contained within the crawler's body. The restrictions on $\dot{x}_1(t)$ become $0 < \dot{x}_1(t) < \varepsilon c$ and, therefore, the force balance reads

$$[\tau_- - \mu_-(\dot{x}_1(t) - \varepsilon c)](1+\varepsilon)\delta - [\tau_+ + \mu_+\dot{x}_1(t)](L - \delta) = 0, \quad (4.16)$$

from which we get

$$\dot{x}_1(t) = \frac{(\tau_- + \mu_-\varepsilon c)(1+\varepsilon)\delta - \tau_+(L-\delta)}{\mu_-(1+\varepsilon)\delta + \mu_+(L-\delta)}. \quad (4.17)$$

This solution is admissible under conditions (4.14), and its time integration in the interval $t \in [\delta/c, L/c)$ yields the second contribution to the displacement as

$$\Delta x_1^b = \frac{\delta(L-\delta)(\tau_- + \mu_-\varepsilon c)(1+\varepsilon)}{\delta(1+\varepsilon)\mu_- c + (L-\delta)\mu_+ c}. \quad (4.18)$$

*Stage c:* $t \in [L/c, (L+\delta)/c)$. During the last time interval the square wave leaves the crawler at its right end. The velocity $\dot{x}_1(t)$ is again restricted to $0 < \dot{x}_1(t) < \varepsilon c$ and the equation for the force balance yields

$$[\tau_- - \mu_-(\dot{x}_1(t) - \varepsilon c)](1+\varepsilon)(L - ct + \delta) - [\tau_+ + \mu_+\dot{x}_1(t)](ct - \delta) = 0, \quad (4.19)$$

from which we get

$$\dot{x}_1(t) = \frac{(\tau_- + \mu_-\varepsilon c)(1+\varepsilon)(L - ct + \delta) - \tau_+(ct - \delta)}{\mu_-(1+\varepsilon)(L - ct + \delta) + \mu_+(ct - \delta)}. \quad (4.20)$$

Also in this case the solution is admissible under conditions (4.14), and integration in the time interval $t \in [L/c, (L+\delta)/c)$ leads to the third contribution of the displacement as

$$\Delta x_1^c = \frac{\delta(1+\varepsilon)(\tau_- + \mu_-\varepsilon c)}{c[(1+\varepsilon)\mu_- - \mu_+]} + \frac{L(1+\varepsilon)(\tau_- + \mu_-\varepsilon c)\mu_+}{c[(1+\varepsilon)\mu_- - \mu_+]^2} \ln\left[\frac{L\mu_+}{(1+\varepsilon)\delta\mu_- + (L-\delta)\mu_+}\right]. \quad (4.21)$$

---

[2]For $(1+\varepsilon)\mu_- = \mu_+$, the displacement (4.15) in the first interval must be replaced by

$$\Delta x_1^a = -\varepsilon\delta + \frac{\delta^2\varepsilon}{2L} + \frac{\delta^2(1+\varepsilon)\tau_-}{2L\mu_+ c}, \quad (4.15^*)$$

and the displacement (4.21) in the third interval by

$$\Delta x_1^c = \frac{\delta^2\varepsilon}{2L} + \frac{\delta^2(1+\varepsilon)\tau_-}{2L\mu_+ c}, \quad (4.21^*)$$

so that the overall displacement after one period, instead of (4.22), becomes

$$\Delta x_1 = -\varepsilon\delta + \frac{\delta^2\varepsilon}{L} + \frac{\delta^2(1+\varepsilon)\tau_-}{L\mu_+ c} + \frac{\delta(L-\delta)(\tau_- + \mu_-\varepsilon c)(1+\varepsilon)}{\delta(1+\varepsilon)\mu_- c + (L-\delta)\mu_+ c}. \quad (4.22^*)$$

Table 2: Admissibility restrictions on the width $\delta$ and on the substrate rheology.

|  | stick-slip crawling | sliding crawling |
|---|---|---|
| $\varepsilon > 0$: | $\delta \leq \dfrac{\tau_+ L}{(\tau_- + \mu_- \varepsilon c)(1+\varepsilon) + \tau_+}, \; \tau_+ \neq 0$ | $\tau_+ = 0, \; \delta < \dfrac{\mu_+ \varepsilon c}{\mu_+ \varepsilon c + \tau_-(1+\varepsilon)} L, \; \mu_+ \neq 0$ |
| $\varepsilon < 0$: | $\delta \leq \dfrac{\tau_- L}{(\tau_+ - \mu_+ \varepsilon c)(1+\varepsilon) + \tau_-}, \; \tau_- \neq 0$ | $\tau_- = 0, \; \delta < \dfrac{\mu_- \varepsilon c}{\mu_- \varepsilon c - \tau_+(1+\varepsilon)} L, \; \mu_- \neq 0$ |

In conclusion, the total net advancement $\Delta x_1$, arising from an extensional wave spanning the crawler's body once, is computed adding equations (4.15), (4.18) and (4.21), leading to

$$\Delta x_1 = \frac{\delta \varepsilon [(1+\varepsilon)\mu_- + \mu_+]}{(1+\varepsilon)\mu_- - \mu_+} + \frac{2\delta(1+\varepsilon)\tau_-}{c[(1+\varepsilon)\mu_- - \mu_+]} + \frac{\delta(L-\delta)(\tau_- + \mu_- \varepsilon c)(1+\varepsilon)}{\delta(1+\varepsilon)\mu_- c + (L-\delta)\mu_+ c} +$$

$$+ \frac{2L(1+\varepsilon)(\tau_- + \mu_- \varepsilon c)\mu_+}{c[(1+\varepsilon)\mu_- - \mu_+]^2} \ln\left[\frac{L\mu_+}{\delta(1+\varepsilon)\mu_- + (L-\delta)\mu_+}\right]. \quad (4.22)$$

The same reasoning holds also in the context of contraction waves ($-1 < \varepsilon < 0$). In that case, $\varepsilon c < \nu(t) < 0$ and the formulae above can still be applied, provided that we replace $\tau_-$ with $-\tau_+$ and $\mu_-$ with $\mu_+$. In particular, we remark that the admissibility conditions (4.14) are replaced in the contractive case by

$$\tau_- = 0 \quad \text{and} \quad \delta < \frac{\mu_- \varepsilon c}{\mu_- \varepsilon c - \tau_+(1+\varepsilon)} L. \quad (4.23)$$

The restrictions on the width $\delta$ and on the substrate rheology deserve particular attention and are summarized in Table 2. In fact, considering the case of an extension (contraction) wave, we may observe that *sliding* crawling requires a vanishing value of $\tau_+$ (resp. $\tau_-$), whereas *stick-slip* crawling is feasible only if $\tau_+ \neq 0$ (resp. $\tau_- \neq 0$). In other words, the two modes of locomotion ('sliding' and 'stick-slip' crawling) are mutually exclusive, in the sense that they are not compatible with the same choice of wave and substrate parameters.

**Newtonian friction** The admissibility conditions (4.14) and (4.23) are quite strict, see also the right column of Table 2. Indeed, sliding crawling by means of both extension and contraction waves is feasible only for a purely Newtonian rheology ($\tau_- = \tau_+ = 0$), and this is also the only case where $\delta$ can freely vary in the interval $(0, L)$. In this context, the net displacement $\Delta x_1$ for an extension wave becomes

$$\Delta_N x_1 = \frac{\delta \varepsilon [(1+\varepsilon)\beta^2 + 1]}{(1+\varepsilon)\beta^2 - 1} + \frac{\delta(L-\delta)(1+\varepsilon)\varepsilon\beta^2}{\delta(1+\varepsilon)\beta^2 + (L-\delta)} +$$

$$+ \frac{2L(1+\varepsilon)\varepsilon\beta^2}{[(1+\varepsilon)\beta^2 - 1]^2} \ln\left[\frac{L}{\delta(1+\varepsilon)\beta^2 + (L-\delta)}\right], \quad (4.24)$$

and the formula for contraction waves can be obtained by replacing $\beta$ with $1/\beta$.

The displacement attained after one stretching cycle is shown in Fig. 7 as a function of $\varepsilon$ for the choice $\delta/L = 0.25$ and for $\beta^2 = \{0.25, 0.5, 1, 2, 4\}$. A lower friction in the direction of wave propagation ($\beta \geq 1$) always leads to a positive displacement, whereas a negative displacement is possible, for sufficiently small values of $\varepsilon$, when friction is lower in the opposite direction ($\beta < 1$). Furthermore, the displacement always tends to $\delta$ as $\varepsilon \to -1$, and tends to $+\infty$ as $\varepsilon \to +\infty$. A decrease of $\beta$ enlarges the range of the values of $\varepsilon$ that produce a negative displacement.

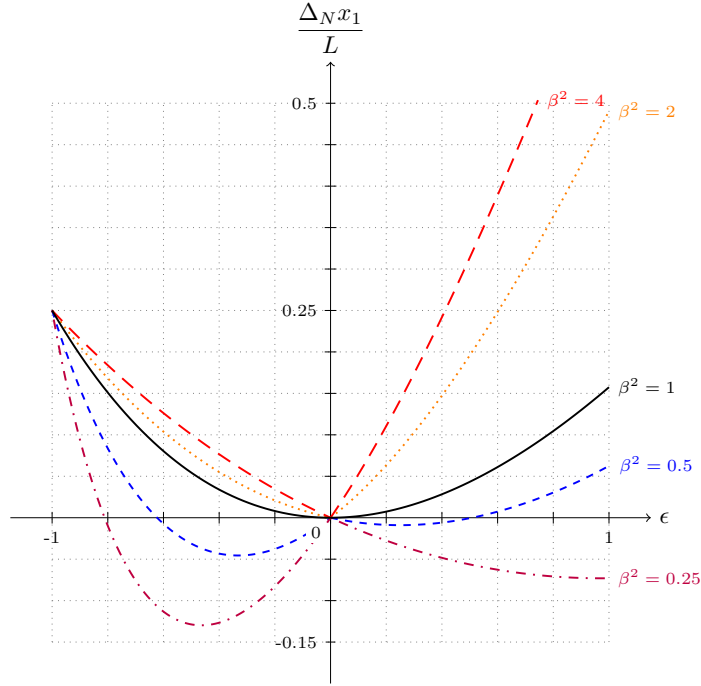

Figure 7: Maximum displacement $\Delta_N x_1/L$ for a sliding crawler in the case of Newtonian friction as a function of $\varepsilon \in (-1, 1)$ for the choice $\delta/L = 0.25$ and for $\beta^2 = \{0.25, 0.5, 1, 2, 4\}$.

## 5 Conclusions and perspectives

In this paper, the motility of one-dimensional crawlers in the presence of directional, *Bingham*-type frictional interaction with a flat substrate has been explored in the regime where inertia effects are negligible. The formulation of the problem within the non-linear framework of large deformations is contained in Section 2, whereas the relevance of such non-linear interactions to crawling motility has been assessed in Section 3 through the analysis of elementary systems, referred to as breathers and constant-length crawlers. In fact, it is shown that 'breathing' modes are effective only when performed on directional substrates. As an example of our results, we call the reader's attention on equations (3.15) and (3.17), showing that the achievable displacement typically equals a fraction of the change of body length in one cycle, and that this fraction tends to one only in the limit case of infinite contrast between the resistance forces in the easy and hard direction. In this ideal limit case, there is no back-sliding, and all the available extension/contraction of the crawler's body is converted into 'useful' displacement, as it is commonly assumed in the classical biological literature [1, 31].

The case of cyclic shape changes arising from localized, extension or contraction waves has been addressed in section 4, and, in this context, two distinct modes of locomotion have been identified (stick-slip and sliding crawling). For both the modes, explicit formulae have been derived for the achievable net displacements (see equations (4.7) and (4.22)), together with the necessary restrictions upon the actuation strategy and the regime of material parameters in which they are feasible (these are summarized in Table 2).

We are presently testing the practical applicability of our findings on actual prototypes [29], which are being manufactured, tested, numerically simulated, and compared with the predictions of effective interaction models such as the one employed in this study.

Future work will consist in exploring the possibility of actually implement these concepts in practical designs exploiting active materials such as nematic elastomers [20, 32], the effect of elastic instabilities [8, 9, 34], or bistable and hysteretic devices [15, 19, 24].

# Acknowledgments

We gratefully acknowledge the support by SISSA through the excellence program NOFYSAS 2012 and by the European Research Council through the ERC Advanced Grant 340685-MicroMotility.

# References

[1] R.M. Alexander, Principles of animal locomotion, Princeton University Press (2006).

[2] F. Alouges, et al., Optimal strokes for low Reynolds number swimmers: an example, J. Nonlinear Sci. 18 (2008) 277-302.

[3] F. Alouges, et al., Optimal strokes for axisymmetric microswimmers, Eur. Phys. J. E 28 (2009) 279-284.

[4] F. Alouges, et al., Optimal swimming of Stokesian robots, Discrete Contin. Dyn. Syst. B 18 (2013) 1189-1215.

[5] F. Alouges, et al., Self-propulsion of slender micro-swimmers by curvature control: N-link swimmers, Int. J. Non-Linear Mech. 56 (2013) 132-141.

[6] M. Arroyo, et al., Reverse engineering the euglenoid movement, Proc. Nat. Acad. Sci. USA 109 (2012) 17874-17879.

[7] M. Arroyo and A. DeSimone, Shape control of active surfaces inspired by the movement of euglenids, J. Mech. Phys. Solids 62 (2014) 99-112.

[8] H.B. Belgacem, et al., Rigorous bounds for the Föppl-von Karman theory of isotropically compressed plates, J. Nonlinear Sci. 10 (2000) 661-683.

[9] D. Bigoni and G. Noselli, Experimental evidence of flutter and divergence instabilities induced by dry friction, J. Mech. Phys. Solids 59 (2011) 2208–2226.

[10] L. Cardamone, et al., Cytoskeletal actin networks in motile cells are critically self-organized systems synchronized by mechanical interactions, Proc. Nat. Acad. Sci. USA 108 (2011)13978-13983.

[11] B. Chan et al., Building a better snail: lubrication and adhesive locomotion, Phys. Fluids 17 (2005) 113101-1:10.

[12] G. Cicconofri and A. DeSimone, A model bristle-bot, submitted for publication (2014).

[13] M. Denny, The role of gastropod pedal mucus in locomotion, Nature 285 (1980) 160-161.

[14] V.S. Deshpande, et al., A bio-chemo-mechanical model for cell contractility, Proc. Nat. Acad. Sci. USA 103 (2006) 14015-14020.

[15] A. DeSimone, Hysteresis and imperfection sensitivity in small ferromagnetic particles, Meccanica 30 (1995) 591-603.

[16] A. DeSimone and A. Tatone, Crawling motility through the analysis of model locomotors: two case studies, Eur. Phys. J. E 35 (2012) 85.

[17] A. DeSimone, et al., Crawlers in viscous environments: linear vs non-linear rheology, Int. J. Non-Linear Mech. 56 (2013) 142-147.

[18] R. Dreyfus, et al., Microscopic artificial swimmers, Nature 437 (2005) 862-865.

[19] L. Fedeli, et al., Metastable equilibria of capillary drops on solid surfaces: a phase field approach, Continuum Mech. and Thermodyn. 23 (2011) 453-471.

[20] A. Fukunaga, et al., Dynamics of electro-opto-mechanical effects in swollen nematic elastomers, Macromolecules 41 (2008) 9389-9396.

[21] D. Fletcher and J. Theriot, An introduction to cell motility for the physical scientist, Phys. Biol. 1 (2004) 1-10.

[22] L. Giomi et al., Swarming, swirling and stasis in sequestered bristle-bots, Proc. R. Soc. A 469 (2013) 20120637.

[23] A. Ghosh and P. Fischer, Controlled propulsion of artificial magnetic nanostructured propellers, Nano Lett. 9 (2009) 2243-2245.

[24] N. Gruenewald et al., A new model for contact angle hysteresis, Netw. Heterog. Media 2 (2007) 211-225.

[25] M.J. Hancock et al., Bioinspired directional surfaces for adhesion, wetting, and transport, Advanced Functional Materials 22 (2012) 2223-2234.

[26] J.H. Lai, et al., The mechanics of the adhesive locomotion of terrestrial gastropods, J. Exp. Biol. 213 (2010) 3920-3933.

[27] L. Mahadevan, et al., Biomimetic ratcheting motion of a soft, slender, sessile gel, Proc. Nat. Acad. Sci. USA 101 (2004) 23-26.

[28] G. Noselli, et al., Discrete one-dimensional crawlers on viscous substrates: achievable net displacements and their energy cost, Mech. Res. Commun. (2013) doi: 10.1016/j.mechrescom.2013.10.023.

[29] G. Noselli and A. DeSimone, Theoretical and experimental study of a model crawler exploiting directional frictional interactions with a substrate, in preparation (2014).

[30] E.M. Purcell, Life at low Reynolds number, Am. J. Phys. 45 (1977) 3-11.

[31] K.J. Quillin, Kinematic scaling of locomotion by hydrostatic animals: ontogeny of peristaltic crawling by the earthworm *Lumbricus terrestris*, J. Exp. Biol. 202 (1999) 661-674.

[32] Y. Sawa et al., Thermally driven giant bending of LCE films with hybrid alignment, Macromolecules, 43 (2010) 4362-4369.

[33] Y. Tanaka, et al., Mechanics of peristaltic locomotion and role of anchoring, J. R. Soc. Interface 9 (2012) 222-233.

[34] D. Zaccaria, et al., Structures buckling under tensile loads, Proc. R. Soc. A 467 (2011) 1686-1700.

[35] L. Zhang, et al., Artificial bacterial flagella: fabrication and magnetic control, Appl. Phys. Lett. 94 (2009) 064107-1:3.